\documentclass{webofc}
\usepackage[varg]{txfonts}   
\usepackage{hyperref}
\usepackage{url}
\usepackage{float}
\usepackage{lineno}
\hypersetup{colorlinks=true,citecolor=blue,urlcolor=blue,linkcolor=blue}
\begin{document}
\title{Quarkonia collectivity in large collision systems with ALICE}
%
%

\author{\firstname{Chi} \lastname{Zhang\inst{1}\fnsep\thanks{\email{chi.zhang@cern.ch}}}on behalf of the ALICE Collaboration.
}

\institute{Université Paris-Saclay, Centre d’Etudes de Saclay (CEA), IRFU, Départment de Physique Nucléaire (DPhN), Saclay, France}

\abstract{Quarkonium production is one of the golden probes to study the quark--gluon plasma (QGP). Among many observables, the measurement of azimuthal anisotropies in quarkonium production sheds light on the collective behavior of heavy-flavor particles in a strongly interacting medium. In particular, the measurements of the elliptic flow ($v_{2}$) of quarkonia in Pb--Pb collisions at the LHC provide us direct evidence of heavy quark thermalization in the QGP. In these proceedings, new results of inclusive $\mathrm{J/\psi}$ elliptic flow measurement in Pb--Pb collisions carried out by the ALICE collaboration in Run 3 using three methods including event-plane, scalar-product and multi-particle correlation (cumulant) will be presented. The method of cumulant will give access to the $\mathrm{J/\psi}$ flow fluctuations at forward rapidity. Alongside the new flow measurements of $\mathrm{J/\psi}$, new results of $\mathrm{\Upsilon}(1S)$ flow measurement at forward rapidity in ALICE Run 3 will be presented as well with comparison to model calculations.}
\maketitle
\section{Introduction}
\label{intro}

Following the recent experimental efforts, one of the most important observations in heavy-ion collisions is the collective flow phenomenon; the idea of anisotropy as a signature of transverse collective flow was first introduced in 1992~\cite{1}. The latter can be evaluated using the Fourier expansion of particle density distribution with the corresponding Fourier harmonics identified as flow coefficients. Among the experimental probes for collective phenomena, quarkonia play a particularly interesting and crucial role. Since heavy quarks are produced in an azimuthally isotropic way in the initial hard processes and they may eventually thermalize in the hot medium, any observed positive elliptic flow signal of quarkonia shall arise from the interaction of constituent heavy quarks with the QGP~\cite{2}. Moreover, the production of quarkonia was found to be strongly affected by the surrounding medium in presence of two competing mechanisms: (1) Suppression with in-medium screening masses for gluons (Debye screening) and thermal masses for quarks, or with the large imaginary part of the in-medium heavy quarkonium potential.~\cite{3} (2) Enhancement from recombination of in-medium (partially) thermalized heavy quarks.~\cite{4,5}\\

With the ALICE detector, quarkonia can be measured either at midrapidity ($|\mathit{y}|<0.9$) via the $\mathit{e}^{+}\mathit{e}^{-}$ decay channel, or at forward rapidity ($2.5 < \mathit{y} < 4$) via the $\mu^{+}\mu^{-}$ decay channel. In these proceedings, only results of quarkonia at forward rapidity will be presented. The main detectors used at midrapidity for the measurement of charged particles as reference particles in the flow correlation studies are: the Inner Tracking System (ITS), used for tracking, vertexing, and the measurement of midrapidity charged-particle multiplicity; the Time Projection Chamber (TPC) used for tracking, particle identification via the measurement of the specific energy loss and also event-plane estimation; the Fast Interaction Triggers (FITs) used for centrality determination and event-plane estimation. At forward rapidity, the measurements of quarkonia are performed with the muon spectrometer, consisting of absorbers, a dipole magnet, tracking and triggering stations, which provide reconstruction and identification of muon tracks. A detailed description of the ALICE apparatus can be found in~\cite{6,7}.

\section{Results}
\label{results}
\begin{figure}[H]
	\centering
	\includegraphics[width=0.45\textwidth]{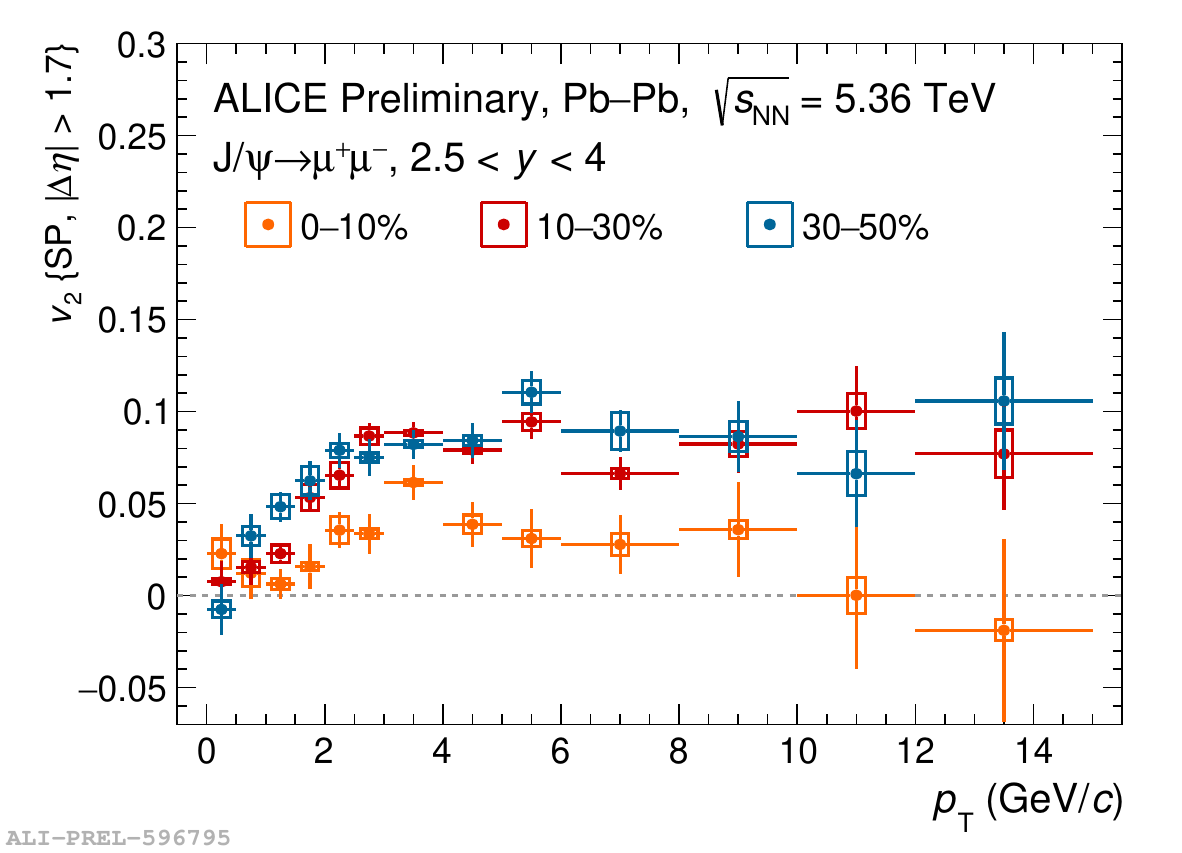}
    \includegraphics[width=0.45\textwidth]{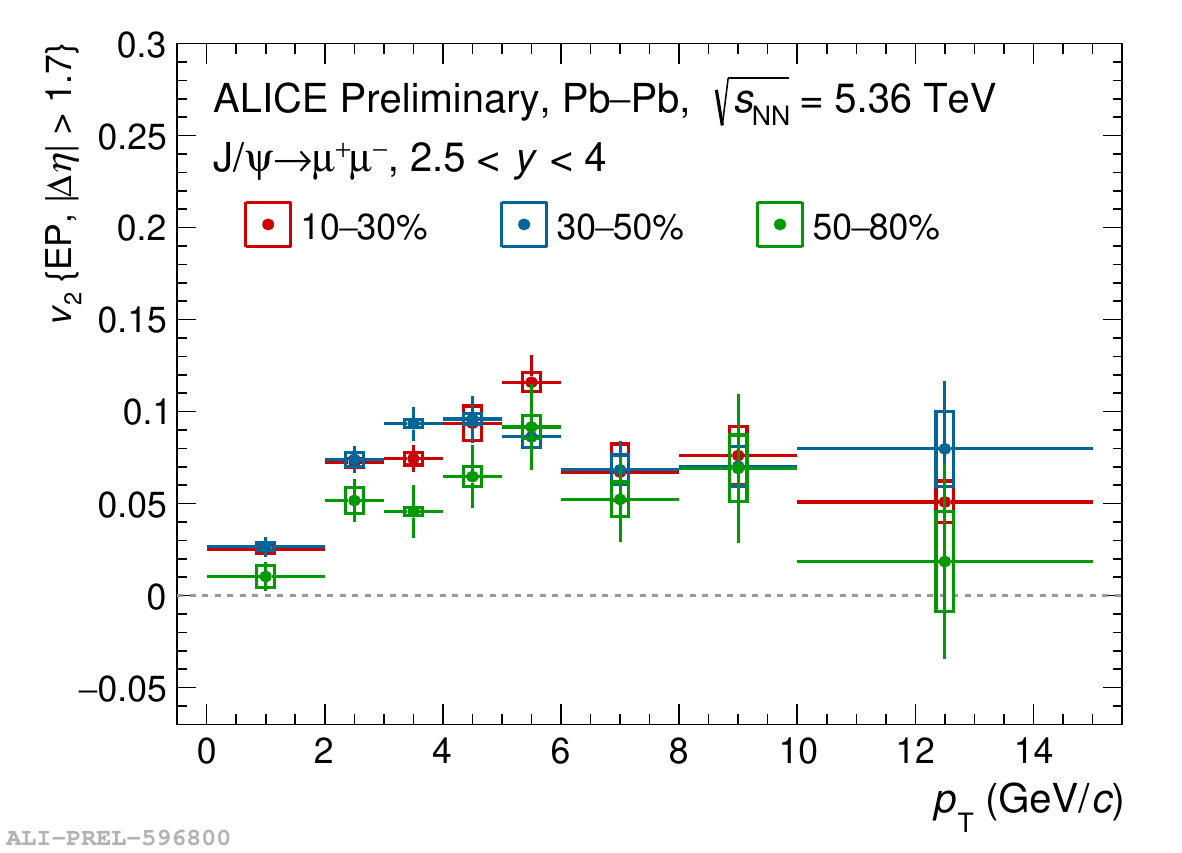}
	\caption{Inclusive $\mathrm{J/\psi}$ $v_{2}$ at forward rapidity in Pb--Pb collisions at $\sqrt{s_{\mathrm{NN}}}=5.36$ TeV as a function of $p_{\mathrm{T}}$. Left: in centrality intervals (0--10\%, 10--30\%, 30--50\%) using the scalar-product method. Right: in centrality intervals (10--30\%, 30--50\%, 50--80\%) using the event-plane method.}
	\label{v2jpsipt}
\end{figure}
\begin{figure}[H]
	\centering
    \includegraphics[width=0.45\textwidth]{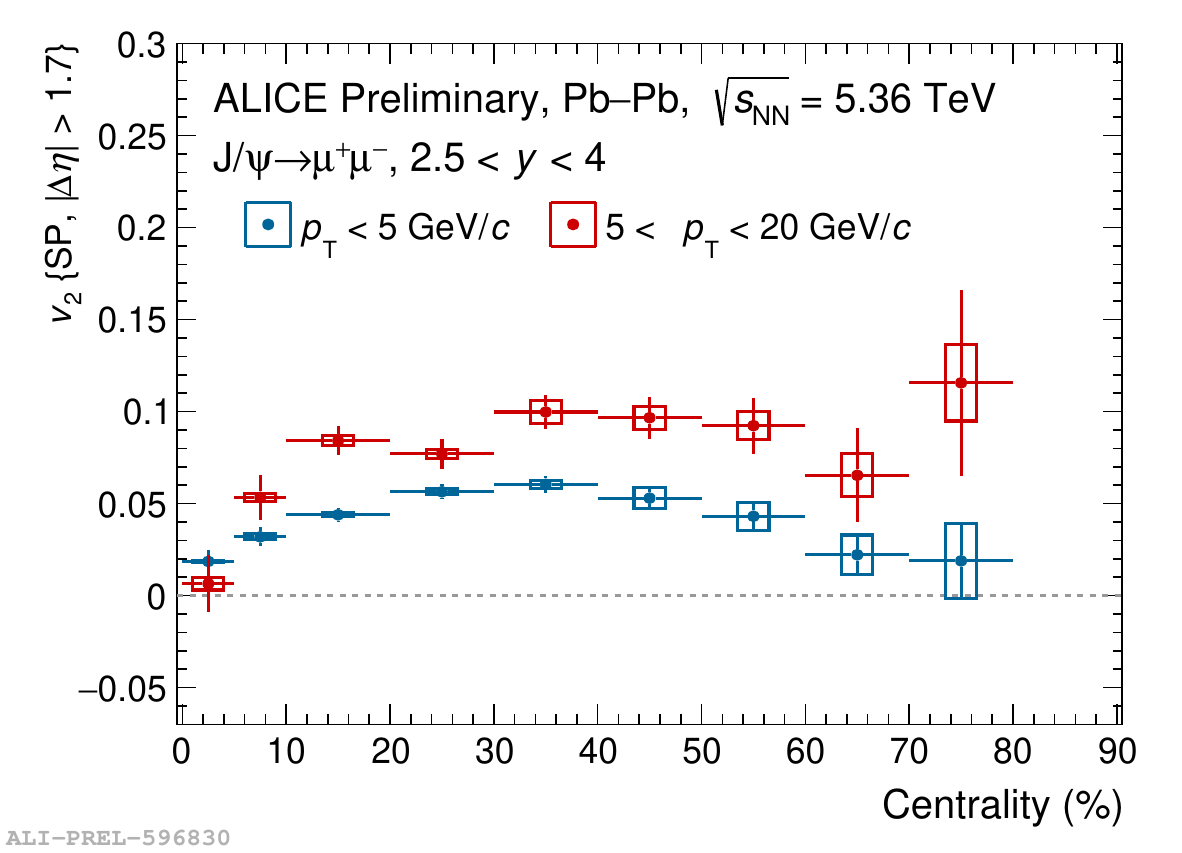}
    \includegraphics[width=0.45\textwidth]{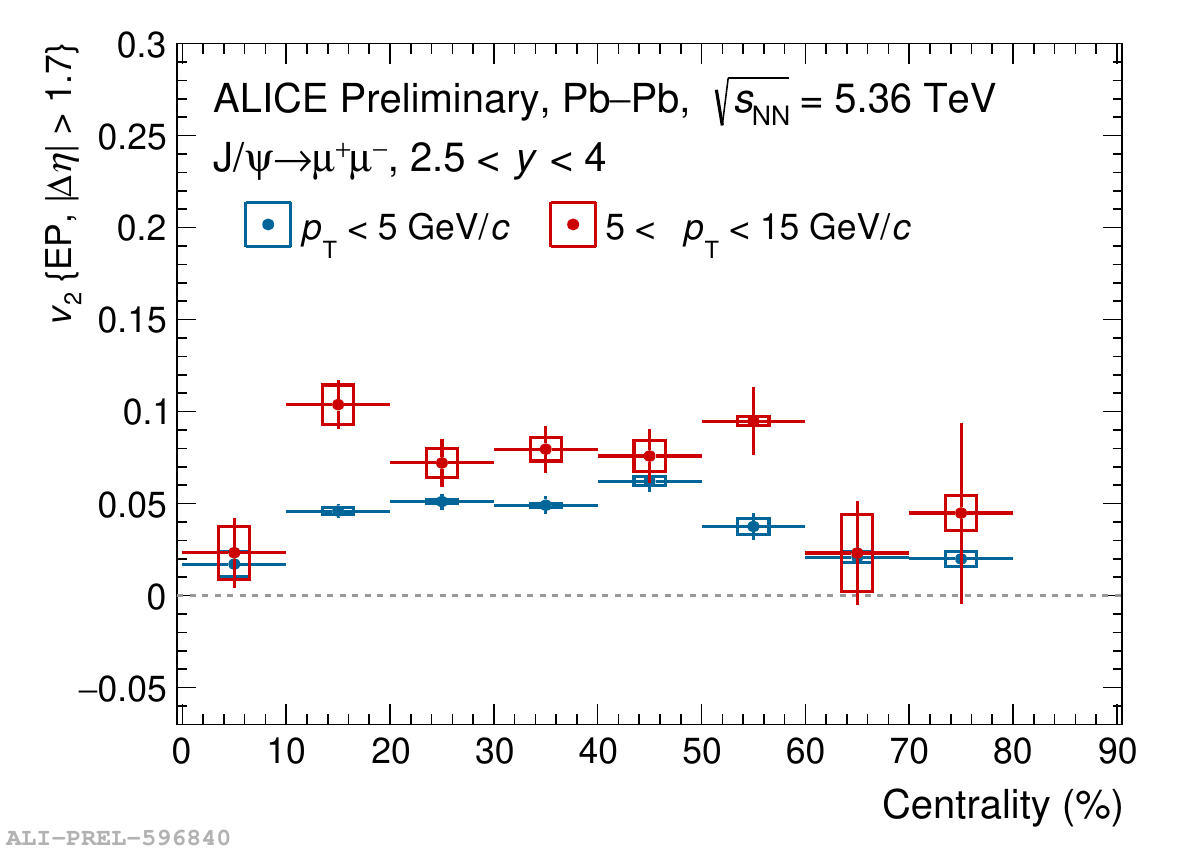}
	\caption{Inclusive $\mathrm{J/\psi}$ $v_{2}$ at forward rapidity in Pb--Pb collisions at $\sqrt{s_{\mathrm{NN}}}=5.36$ TeV as a function of centrality in different $p_{\mathrm{T}}$ intervals (0--5 GeV/$c$, 5--20 (15) GeV/$c$). Left: the scalar-product method. Right: the event-plane method.}
	\label{v2jpsicent}
\end{figure}
In these proceedings, the ALICE Run 3 preliminary results of inclusive $\mathrm{J/\psi}$ $v_{2}$ with three different methods including event-plane~\cite{8}, scalar-product~\cite{9} and multi-particle correlation (cumulant)~\cite{10} are presented. In Fig.~\ref{v2jpsipt}, the inclusive $\mathrm{J/\psi}$ $v_{2}$ at forward rapidity in Pb--Pb collisions at $\sqrt{s_{\mathrm{NN}}}=5.36$ TeV as a function of $p_{\mathrm{T}}$ with event-plane and scalar-product methods are presented, within different centrality intervals. The results from both methods within 10--30\% and 30--50\% centrality intervals agree within uncertainties. The observed rising trend from low to intermediate $p_{\mathrm{T}}$ regions is well consistent with charm thermalization description; the trend toward high $p_{\mathrm{T}}$ follows the path-length dependent energy-loss effect~\cite{11} and a significant contribution from the hydrodynamic flow. The higher amount of data from Run 3 allows the use of finer $p_{\mathrm{T}}$ binning and hence more precise results as a function of $p_{\mathrm{T}}$. The corresponding centrality-differential measurements are shown in Fig.~\ref{v2jpsicent} using event-plane and scalar-product methods with improved granularity. In the low $p_{\mathrm{T}}$ region, where the regeneration component is expected to dominate, $v_{2}$ increases from 0 in the most central collisions, reaches a maximum around 30--40\% midcentral collisions, then decreases toward peripheral collisions; in the high $p_{\mathrm{T}}$ region, $v_{2}$ shows a increasing trend from the most central to peripheral collisions.

\begin{figure}[H]
	\centering
	\includegraphics[width=0.45\textwidth]{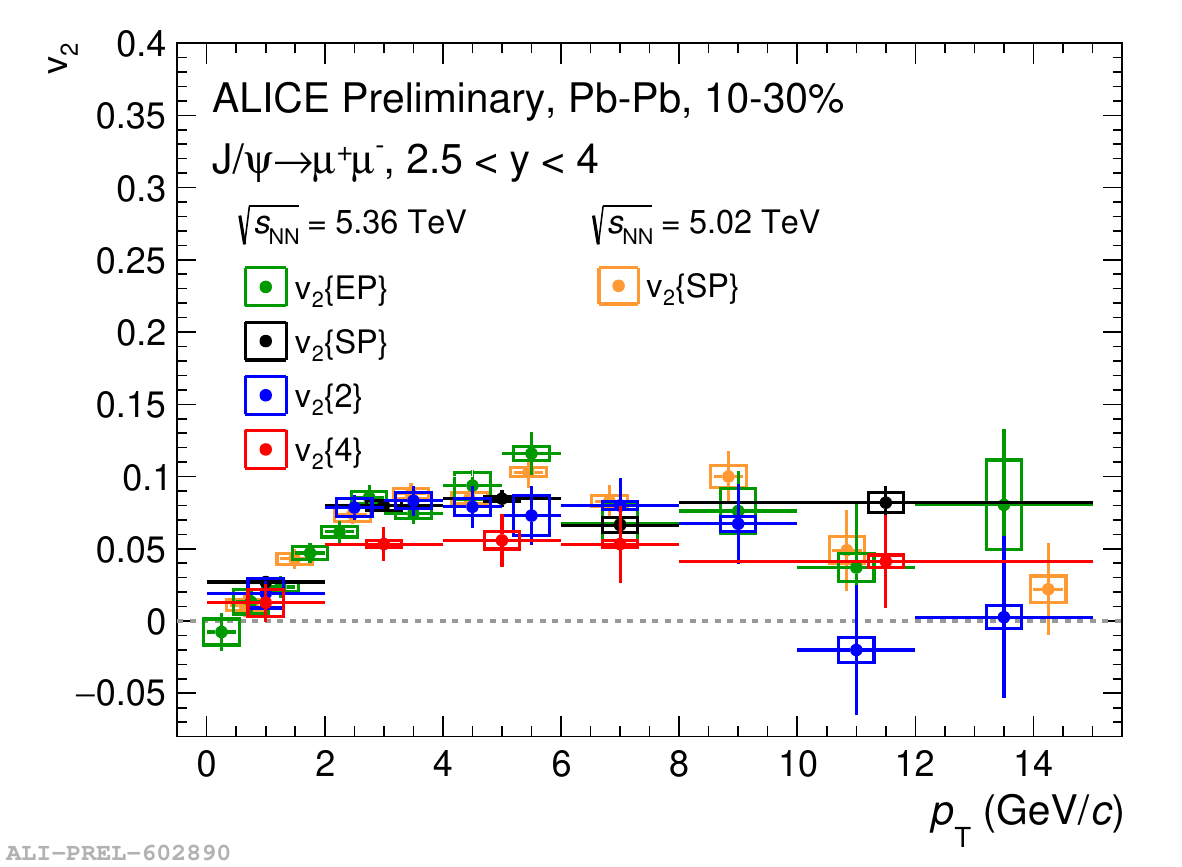}
    \includegraphics[width=0.48\textwidth]{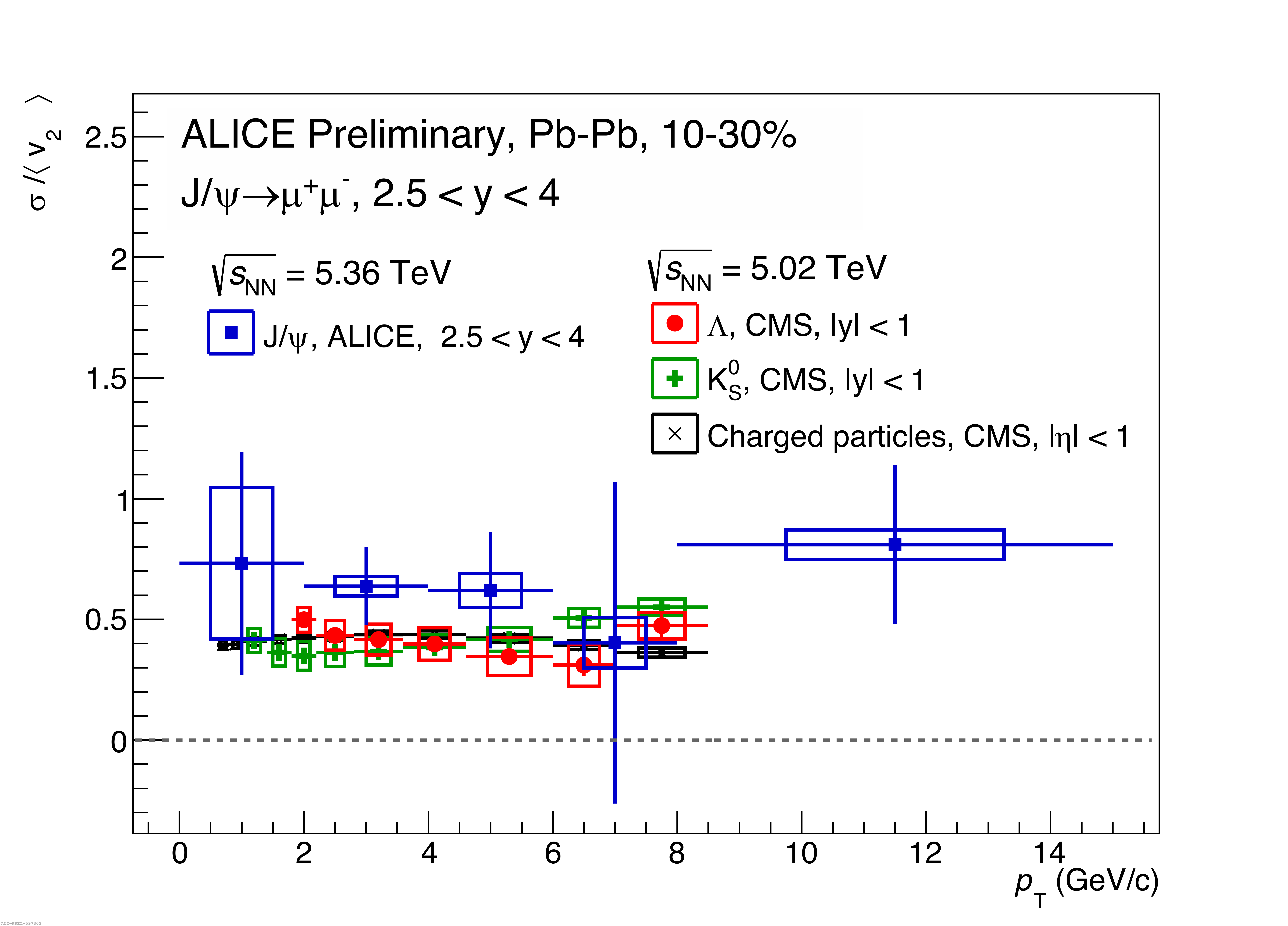}
	\caption{Left: inclusive $\mathrm{J/\psi}$ $v_{2}$ at forward rapidity in Pb--Pb collisions at $\sqrt{s_{\mathrm{NN}}}=5.36$ TeV as a function of $p_{\mathrm{T}}$ within 10--30\% centrality using the method of cumulant compared to event-plane, scalar product methods, and to the Run 2 results~\cite{12}. Right: the $\mathrm{J/\psi}$ flow fluctuation at forward rapidity in Pb--Pb collisions at $\sqrt{s_{\mathrm{NN}}}=5.36$ TeV as a function of $p_{\mathrm{T}}$ within 10--30\% centrality compared to the CMS measurements of light-flavor particles~\cite{13}.}
	\label{v2jpsicumulant}
\end{figure}

The larger dataset collected by ALICE in Run 3 makes possible the measurement of the $\mathrm{J/\psi}$ $v_{2}$ using higher-order correlations which will suppress the non-flow~\footnote{From jets, resonance decays etc.} contribution. The results of inclusive $\mathrm{J/\psi}$ $v_{2}$ at forward rapidity in Pb--Pb collisions using the method of cumulant are presented in the left panel of Fig.~\ref{v2jpsicumulant}, where the results of $v_{2}\{2\}$ and $v_{2}\{4\}$ as a function of $p_{\mathrm{T}}$ are found to be consistent with $v_{2}$ from the scalar-product method. The $\mathrm{J/\psi}$ flow fluctuation shown in the right panel of Fig.~\ref{v2jpsicumulant} can be estimated as $\frac{\sigma_{v_{2}}}{<v_2>}\simeq\sqrt{\frac{v_{2}\{\mathrm{SP}\}^2-v_{2}\{4\}^2}{v_{2}\{\mathrm{SP}\}^2+v_{2}\{4\}^2}}$. In case the ratio does not depend on $p_{\mathrm{T}}$, then any observed non-zero fluctuation ratio should arise from the variation in initial geometry. In our result, a small hint of non-zero fluctuation ratio is found across a large range of $p_{\mathrm{T}}$, and our result is consistent with the CMS measurements of light-flavor particles~\cite{13}.\\

\begin{figure}[H]
	\centering
	\includegraphics[width=0.45\textwidth]{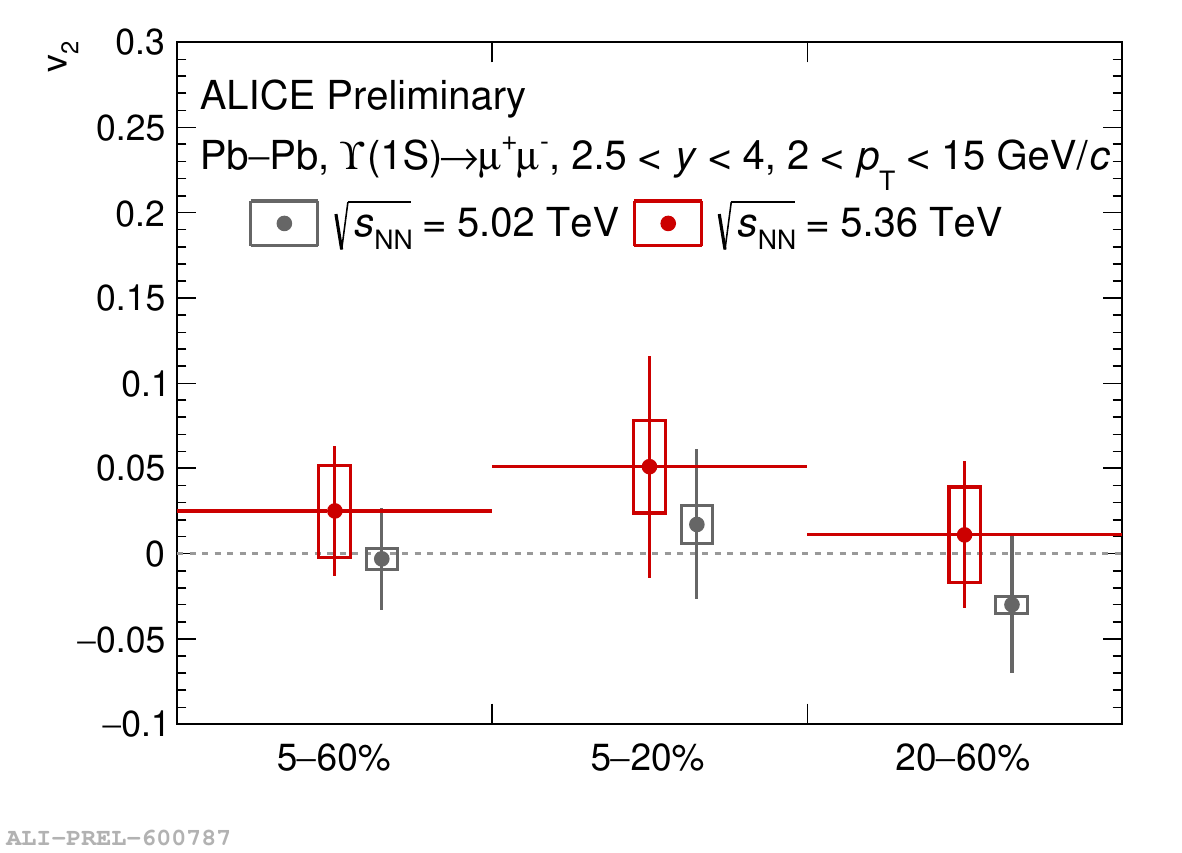}
    \includegraphics[width=0.45\textwidth]{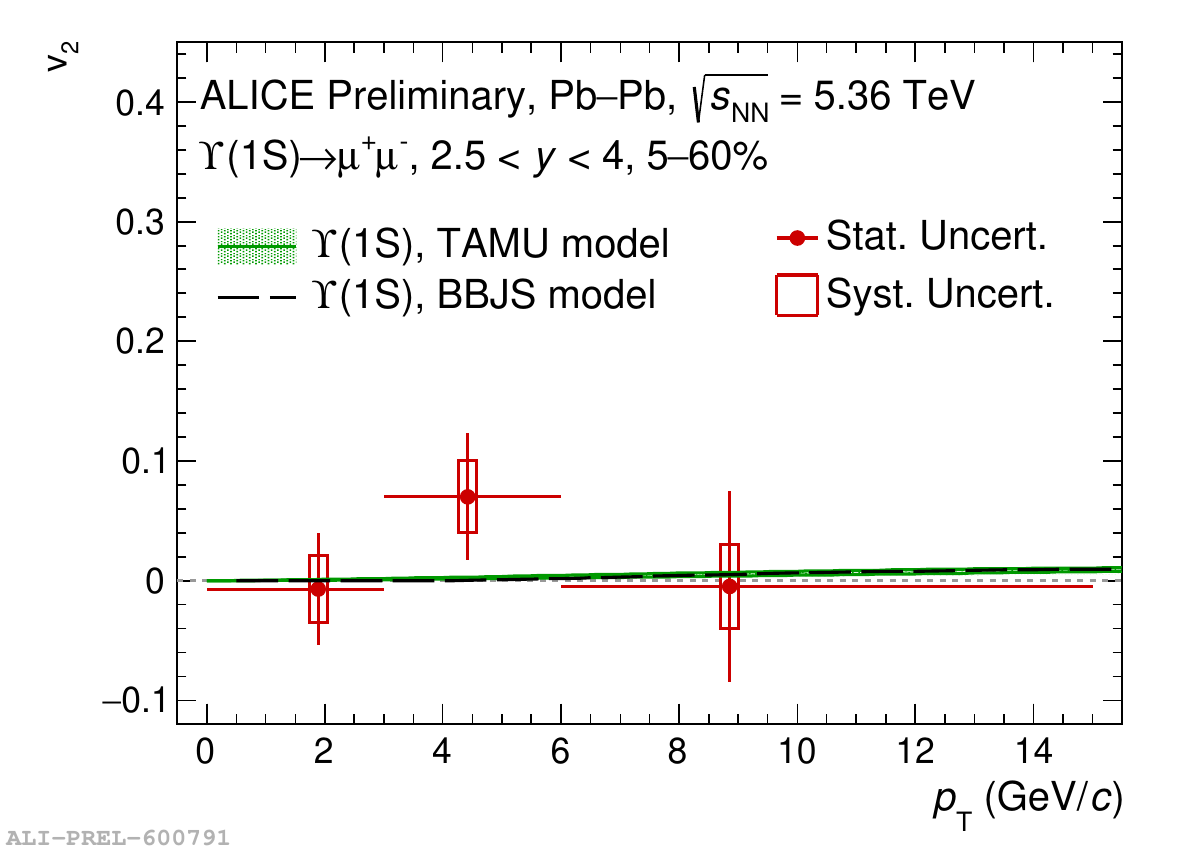}
	\caption{Inclusive $\mathrm{\Upsilon}(1S)$ $v_{2}$ at forward rapidity in Pb--Pb collisions at $\sqrt{s_{\mathrm{NN}}}=5.36$ TeV. Left: $p_{\mathrm{T}}$-integrated results within different centrality intervals compared to the Run 2 results~\cite{14}. Right: results as a function of $p_{\mathrm{T}}$ within 5--60\% centrality compared to calculations from TAMU~\cite{15} and BBJS~\cite{16} models.}
	\label{v2upsilon}
\end{figure}

In addition, the Run 3 preliminary results of inclusive $\mathrm{\Upsilon}(1S)$ $v_{2}$ at forward rapidity in Pb--Pb collisions are presented in Fig.~\ref{v2upsilon}. The $p_{\mathrm{T}}$-integrated results within different centrality intervals are compared to the Run 2 results~\cite{14}, both results are compatible with 0 within uncertainties. Meanwhile, the $p_{\mathrm{T}}$-differential result shows no significant positive elliptic flow signal within current uncertainties. The result was compared with model calculations including: the TAMU model~\cite{15}, which implements a regeneration component from the recombination of (partially) thermalized bottom quarks; and the BBJS model~\cite{16} considering only the path-length dependent dissociation of initially created in-medium bottomonia. However, the latter comparison makes no discrimination yet considering current uncertainties.

\section{Summary}
In these proceedings, the ALICE preliminary results on investigating quarkonium collectivity in large collision systems with the measurement of elliptic flow using Run 3 data are presented. The new results of inclusive $\mathrm{J/\psi}$ $v_{2}$ at forward rapidity in Pb--Pb collisions are presented with higher amount of data and finer binning. Results show smoother evolution of elliptic flow coefficient as a function of $p_{\mathrm{T}}$ and centrality. Moreover, the result of $\mathrm{J/\psi}$ flow fluctuation using $v_{2}\{4\}$ from the method of cumulant shows a small hint of positive signal across a large range of $p_{\mathrm{T}}$. The new results of inclusive $\mathrm{\Upsilon}(1S)$ $v_{2}$ at forward rapidity in Pb--Pb collisions are presented as well with no evidence of positive flow signal regarding the current uncertainties.

%
%
%

\end{document}